\documentclass[onecolumn,aps,prd,preprintnumbers,showpacs,superscriptaddress,nofootinbib,amsmath,amssymb,floats,floatfix,showkeys,notitlepage,longbibliography]{revtex4-1}

\usepackage{comment}
\usepackage{graphicx}
\usepackage{subfigure}
\usepackage{palatino}
\usepackage[commandnameprefix=always]{changes}
\usepackage{hyperref}
\hypersetup{colorlinks=true,linkcolor=blue,urlcolor=blue,citecolor=blue}
\usepackage[toc,page]{appendix}
\usepackage[normalem]{ulem}

\usepackage{orcidlink}
\usepackage{lipsum}
\usepackage{graphicx}
\usepackage{subfigure}
\usepackage{palatino}
\usepackage{sans}
\usepackage{adjustbox}
\usepackage{latexsym}
\usepackage{amsmath}
\usepackage{amssymb}
\usepackage{amsfonts}
\usepackage{dcolumn}
\usepackage{bm}
\usepackage{tikz}
\usepackage{bigints}
\usepackage{array,tabularx,multirow,booktabs}
\usepackage[tracking=true]{microtype}
\SetTracking{}{500}
\SetTracking{encoding={*}, shape=sc}{40}
\UseRawInputEncoding 
\allowdisplaybreaks
\usepackage{adjustbox}
\usepackage{latexsym}
\usepackage{amsmath}
\usepackage{amssymb}
\usepackage{amsfonts}
\usepackage{dcolumn}
\usepackage{bm}
\usepackage{tikz}
\usepackage{bigints}
\usepackage{array,tabularx,multirow,booktabs}
\usepackage[tracking=true]{microtype}
\usepackage{color}

\def\1o2{{1\over2}}

\newcommand{\ba}{\begin{eqnarray}}
\newcommand{\ea}{\end{eqnarray}}

\def\ben{\begin{equation}}
\def\een{\end{equation}}
\def\bea{\begin{eqnarray}}
\def\eea{\end{eqnarray}}
\def\be{\begin{equation}}
\def\ee{\end{equation}}

\def\ben{\begin{equation}}
\def\een{\end{equation}}
\def\bea{\begin{eqnarray}}
\def\eea{\end{eqnarray}}

\begin{document} 

\title{Analyzing the Influence of Geometrical Deformation on Photon Sphere and Shadow Radius: A New Analytical Approach -Stationary, and Axisymmetric Spacetime}

\author{Vitalii Vertogradov}
\email{vdvertogradov@gmail.com}
\affiliation{Physics department, Herzen state Pedagogical University of Russia,
48 Moika Emb., Saint Petersburg 191186, Russia} 
\affiliation{SPB branch of SAO RAS, 65 Pulkovskoe Rd, Saint Petersburg
196140, Russia}

\author{Ali \"Ovg\"un
}
\email{ali.ovgun@emu.edu.tr}
\affiliation{Physics Department, Eastern Mediterranean
University, Famagusta, 99628 North Cyprus, via Mersin 10, Turkiye}

\author{Reggie C. Pantig}
\email{rcpantig@mapua.edu.ph}
\affiliation{Physics Department, Map\'ua University, 658 Muralla St., Intramuros, Manila 1002, Philippines}

\begin{abstract}
Black hole shadows and photon spheres offer valuable tools for investigating black hole properties. Recent observations by the Event Horizon Telescope Collaboration have confirmed the existence of rotating black holes. Black hole parameters influence the observed shadow size. This paper explores the impact of geometrical deformations on black hole shadow size using gravitational decoupling applied to axially-symmetric spacetime. We find the results are more complex than the spherically-symmetric case. We compare shadows in well-known models with those of an Kerr black hole. Our approach suggests that the influence of an accretion disc on the observed shadow shape can be accurately described despite negligible impact on the black hole geometry itself. 
\end{abstract}

\date{\today}

\keywords{Rotating Black hole; Shadow; Photon sphere; Gravitational decoupling; Deformation}

\pacs{95.30.Sf, 04.70.-s, 97.60.Lf, 04.50.Kd }

\maketitle

\section{Introduction}
Black hole shadow is one of the observable quantities of a black hole. The spacetime is curved so strong near these objects that light can move along circular orbit which forms a black spot on the observer's sky $\sim 2.5$ times bigger than Euclidean size\footnote{Here, under Euclidean size we mean light propagation along straight lines. In this case the visible black spot is the size of the event horizon .}. Recent observation by Event horizon Collaboration Telescope~\cite{bib:eht1, bib:eht2} revealed the black hole in the centre of galaxies M87 and Milky Way. There are  many papers exploring properties of static black holes~\cite{bib:tsupko_plasma,bib:tsupko_deflection,bib:tsupko_lensing,bib:tsupko_kerr, bib:klaudel, bib:hod,Virbhadra:2024pru,Adler:2022qtb,Virbhadra:2022iiy,Virbhadra:2007kw,Virbhadra:1999nm,Virbhadra:1998dy, bib:khoo, bib:decaniny, bib:shoom, bib:sederbaum, bib:jokhannsen, bib:teo,Okyay:2021nnh,Pantig:2022gih,Kuang:2022xjp, Zakharov:2023lib,Zakharov:2021gbg,Qiao:2022hfv,Hod:2020pim,Lu:2019zxb,Paithankar:2023ofw,Aratore:2024bro,Tsupko:2022kwi,bib:rezzolla, bib:rezzolla2, bib:expanding, bib:vaidya1, bib:vaidya2, bib:ver_shadow, bib:tsupko_first, bib:understanding, bib:joshi_shadow,Tsukamoto:2014tja,Tsukamoto:2017fxq,Gomez:2024ack,Adair:2020vso,Ghosh:2023kge}. (See also a recent review~\cite{bib:tsupko_review}). The black hole shadow can serve as a cosmological ruler~\cite{bib:tsupko_ruller, bib:vanozi_ruller} and as a means to test various black hole models~\cite{bib:vanozi}. Real astrophysical black hole should be surrounded by an accretion disc which has an impact on visible size of a shadow. Indeed, one approach involves considering a black hole surrounded by plasma and examining how light propagation within it affects the angular size of the shadow. Alternatively, the black hole-accretion disk system can be treated as a solution of the Einstein field equations, allowing for the analysis of light trajectories within the obtained spacetime.

One can describe a black hole surrounded by an accretion disc by using gravitational decoupling method~\cite{bib:gd1, bib:gd2, Contreras:2021yxe}. It is a novel tool which allows, under some conditions, consider a solution of Einstein field equation as a straight superposition of two solutions. One can consider the deformation of $g_{rr}$ component of the metrec (minimal geometrical deformation) or one can deform both $g_{rr}$ and $g_{tt}$ metric components (extended gravitational decoupling). Using gravitational decoupling, several new solutions of the Einstein field equations have been discovered, depicting deformed versions of well-known black hole solutions~\cite{bib:bh1, bib:bh2, bib:ovalle_regular, bib:max, bib:max_regular, bib:kiselev, bib:rotating, bib:varmhole}. These deformations arise due to the presence of an additional anisotropic matter field surrounding the black hole. In recent paper~\cite{Vertogradov:2024qpf}, the effect of minimal geometrical deformations, which can be caused by the extra matter field surrounding a black hole, on the radius of a photon sphere and visible angular size has been explored. One more important fact regarding real astrophysical black holes is that they should be rotational ones. The comparison of obtained images with Kerr spacetime surrounded by an accretion disc revealed angular momentum of a black hole around $a\sim 0.7$~\cite{bib:doc}.

In this paper, we investigate the influence of minimal geometrical deformations on apparent shape of a shadow in a general axially-symmetric black hole. As opposed to spherical-symmetric black hole, which casts a shadow as a circle, in rotational case the shadow is deformed and flattened on one side. The apparent shape of a shadow depends on the angle of observation. In order to describe the shadow of rotational black hole, one needs to introduce two impact parameters $\xi$ and $\eta$. If one considers the geometrical deformations of the seed spacetime then the impact parameters and apparent shape changes. Here, we describe the influence of these deformations on the apparent shape and compare results with a shadow of a seed spacetime.

The paper organized as follows: In sec.II we describe the method of calculation of a shadow in general axially-symmetric spacetime. In sec. III we elaborate the influence of minimal geometrical deformations on the apparent shape of a black hole shadow. We give several examples in sec. IV and discuss obtained results in sec. V.
The system of units $G=c=1$ and signature $-+++$ will be used throughout the paper.

\section{The influence of deformation on shadow of axially symmetric and rotating black holes}

 Without loss of generality, we can assume
\cite{Vertogradov:2024qpf}
\begin{widetext}
\begin{eqnarray} \label{eq:metric}
ds^2=-\left(1-\frac{M(r)}{r}\right)e^{\alpha \gamma(r)}dt^2+\left(1-\frac{M(r)}{r}\right)^{-1} e^{-\alpha \gamma(r)}dr^2+r^2d\Omega^2,
\end{eqnarray}

\end{widetext}
and for the axially symmetric case one can write it in this form \cite{Contreras:2021yxe}:
\begin{equation} \label{kerrex2}
   ds^{2} = \left[1-\frac{2\,r\,{m}(r)}{\rho^2}\right] dt^{2} +\frac{4\, a\, r\,{m}(r) \sin^{2}\theta}{\rho^{2}} \,dt\,d\phi -\frac{\rho^{2}}{\Delta}\,dr^{2} -\rho^{2}d\theta^{2}-\frac{\Sigma \sin^{2}\theta}{\rho^{2}}d\phi^{2},
\end{equation}
where the expressions for $\rho$, $\Sigma$ and $\Delta$
are
\begin{eqnarray}
m(r)=\frac{1}{2} \left[e^{\alpha  \gamma (r)}(2M-r)+r\right],
\end{eqnarray}
with
\begin{eqnarray}
\rho^2
&=&
r^2+a^{2}\cos^{2}\theta,
\label{f0}
\\
\tilde{\Delta}
& = &
r^2-2\,r\,m(r)
+a^{2},
\label{f2}
\\
\Sigma
& = &
\left(r^{2}+a^{2}\right)^{2}
-a^{2}\, \Delta\, \sin^{2} \theta,
\label{f3}
\end{eqnarray}
and 
\begin{equation}
\label{a}
a\,=\,J/M\ ,
\end{equation}
where $J$ is the angular momentum and $M$ the total mass of the system.
Note that the line-element~\eqref{kerrex2} reduces to the Kerr solution
when the metric function $m=M$.
Moreover, when $a=0$, we obtain the Schwarzschild-like metric.

We briefly review how to study null geodesics in a rotating space-time
like the one in Eq.~\eqref{kerrex2}, 
and find the celestial coordinates describing the BH shadow.
The Hamilton-Jacobi
equation
\begin{equation}
\label{jac}
\frac{\partial S}{\partial\lambda}
=
\frac{1}{2}\,g^{\mu\nu}\,\partial_{\mu}S\,\partial_{\nu}S
\ ,
\end{equation}
where $\lambda$ is a parameter along the curve and $S$ the Jacobi action.
Given the symmetries of the space-time~\eqref{kerrex2}, Eq.~(\ref{jac}) is separable
and one has
\begin{equation}
\label{separable}
S
=
-E\,t+\Phi\,\phi+S_{r}(r)+S_{\theta}(\theta)
\ ,
\end{equation}
with $E$ and $\Phi$ being the conserved energy and angular momentum, respectively.
Replacing~\eqref{separable} in Eq.~\eqref{jac}, we obtain
\begin{eqnarray}
S_{r}
& = & 
\int\frac{\sqrt{R(r)}}{\Delta}\,dr
\nonumber
\\
\\
S_{\theta}
& = &
\int\sqrt{\Theta(\theta)}\,d\theta
\ ,
\nonumber
\end{eqnarray}
where
\begin{eqnarray}
R
&=&
\left[
(r^{2}+a^{2})\,E
-a\,\Phi
\right]^{2}
-
\Delta
\left[
Q+
(\Phi-a\,E)^{2}
\right]
\nonumber
\\
\\
\Theta
&=&
Q
-
(\Phi^{2}\,\csc^{2}\theta-a^{2}\,E^{2})\,\cos^{2}\theta
\ ,
\nonumber
\end{eqnarray}
with $Q$ the Carter constant.
\par
The (unstable) circular photon orbits are determined by $R=R'=0$, namely
\begin{eqnarray}
&&
\left(a^2-a\, \xi +r^2\right)^2
-
\left(a^2+r^2 F\right)
\left[
(a-\xi )^2+\eta
\right] 
= 0
\nonumber
\\
\\
&&
4 \left(a^2-a \,\xi +r^2\right)
-
\left[
(a-\xi )^2+\eta \right]
\left(r\, F'+2\, F\right)
= 0
\ ,
\nonumber
\end{eqnarray}
where $\xi=\Phi/E$ and $\eta=Q/E^{2}$ are the impact parameters.
Accordingly, 
\begin{eqnarray} \label{eq:impact}
\xi
&=&
a
+\frac{r^2}{a}
-\frac{4 \left(a^2+r^2\, F\right)}{a\left(r\, F'+2\, F\right)}
\ ,
\nonumber
\\
\eta
&=&
\frac{r^2 \left[r^2+2\,a \left(a-\xi \right)-\left(a-\xi \right)^2 F\right]}{a^2+r^2\, F}
\ ,
\end{eqnarray}
where $r$ is the radius of the unstable photon orbit.




\par
The apparent shape of the shadow is finally described by
the celestial coordinates
\begin{eqnarray}
\sigma
&\equiv&
\lim\limits_{r_{0}\to \infty}
\left(-r_{0}^{2}\,\sin\theta_{0}\frac{d\phi}{dr}\bigg|_{(r_{0,\theta_{0}})}\right)
\nonumber
\\
&=&
-\frac{\xi }{\sin\theta_{0}}
\label{sigmaf}
\end{eqnarray}
and
\begin{eqnarray}
\beta
&\equiv&
\lim\limits_{r_{0}\to\infty}
\left(r_{0}^{2}\,\frac{d\theta}{dr}\bigg|_{(r_{0},\theta_{0})}\right)
\nonumber
\\
&=&
\sqrt{\eta -\xi ^2 \,\cot ^2\theta_{0}+a^2\, \cos ^2\theta_{0}}
\ ,
\label{betaf}
\end{eqnarray}
where $(r_{0},\theta_{0})$ are the coordinates of the observer.

\section{The influence of geometrical deformation on apparent shape of a shadow}

The dimensionless parameter $\alpha$ is supposed to be small $\alpha \ll 1$, so one can consider its influence on the apparent shape of a shadow which depends on impact parameters $\xi$ and $\eta$ \eqref{eq:impact} by assuming
\begin{eqnarray} \label{eq:impact_assum}
\xi&=&\xi_0+\alpha \xi_1,\nonumber \\
\eta&=&\eta_0+\alpha \eta_1,
\end{eqnarray}
here $\xi_0$ and $\eta_0$ are impact parameters along null geodesics in seed spacetime, $\xi_1$ and $\eta_1$ are small changes caused by presence of an extra matter field. We can write $\xi$ and $\eta$ as functions of $\alpha$ and expand with respect to coupling constant. By notesing that $\xi(0)=\xi_0$ and $\eta(0)=\eta_0$, substituting \eqref{eq:impact_assum} into \eqref{eq:impact} and expanding, we find $\xi_1$ and $\eta_1$ as
\begin{widetext}
\begin{eqnarray} \label{eq:impact_result}
\xi_1&=&\frac{r \left(r (r-2 M) \left(a^2+r (r-2 M)\right) \gamma '(r)+2 a^2 (r-M) \gamma (r)\right)}{a (M-r)^2},\nonumber \\
\eta_1&=&-\frac{2 r^3 \left(r (r-3 M) (r-2 M) \left(a^2+r (r-2 M)\right) \gamma '(r)+2 a^2 M (M-r) \gamma (r)\right)}{a^2 (M-r)^3}.
\end{eqnarray}
\end{widetext}
 where for the seed Schwarzschild spacetime impact parameters are:

\begin{eqnarray} \label{xi0ali}
\xi_0=\frac{a^2 (M+r)+r^2 (r-3 M)}{a (M-r)}
\end{eqnarray}

and
\begin{eqnarray} \label{eta0ali}
\eta_0=-\frac{r^3 \left(r (r-3 M)^2-4 a^2 M\right)}{a^2 (M-r)^2}.
\end{eqnarray}

As one can notice to estimate the influence of geometrical deformation, one should know the impact parameters in seed spacetime and the value of $\gamma$ and $\gamma'$.By substituting \eqref{eq:impact_assum}into \eqref{sigmaf} and \eqref{betaf} with $\xi_1$ and $\eta_1$ from \eqref{eq:impact_result}, we find the apparent shape of deformed axially-symmetric black hole
\begin{eqnarray} \label{eq:sigma_beta}
\sigma &=&-\frac{\xi_0+\alpha \xi_1}{\sin \theta_0},\nonumber \\
\beta&=&\sqrt{\eta_0+\alpha \eta_1-\xi_0(\xi_0+2\alpha \xi_1)\cot ^2\theta_0+a^2\cos^2\theta_0}.
\end{eqnarray}

The celestial coordinates $\sigma$ and $\beta$ show the apparent perpendicular distances of the image
around the black hole. At  $\theta_0 = \pi/2$, the celestial coordinates become 

\begin{eqnarray} \label{eq:sigma_beta2}
\sigma &=&
-\xi_0-\alpha  \xi_1,\nonumber \\
\beta&=&\sqrt{\eta_0+\alpha  \eta_1}.
\end{eqnarray}

 The photonsphere radius $r_{\rm ph}$ can then be determined by solving $\beta=0$ for r. After substituting all parameters we find
 \begin{widetext}
\begin{eqnarray} \label{eq:sigma_beta3}
\sigma &=&\frac{\alpha  r^2 (2 M-r) \left(a^2+r (r-2 M)\right) \gamma '(r)-(M-r) \left(a^2 (M+r)-2 \alpha  a^2 r \gamma (r)+r^2 (r-3 M)\right)}{a (M-r)^2}
,\nonumber \\
\beta&=&\sqrt{\frac{r^3 \left((r-M) \left(4 \alpha  a^2 M \gamma (r)-4 a^2 M+r (r-3 M)^2\right)-2 \alpha  r (r-3 M) (r-2 M) \left(a^2+r (r-2 M)\right) \gamma '(r)\right)}{a^2 (M-r)^3}}.
\end{eqnarray}
\end{widetext}

Note that  one can also define \cite{Feng:2019zzn}

\begin{equation} \label{shadowradius}
R_{\mathrm{sh}}=\frac{1}{2}\left[\sigma\left(r_{\mathrm{ph}}^{+}\right)-\sigma\left(r_{\mathrm{ph}}^{-}\right)\right].
\end{equation}

\subsection{Example 1: Hairy Kerr spacetime by gravitational decoupling}
\begin{figure}
    \centering
\label{hairfig}
\includegraphics[width=0.48\textwidth]{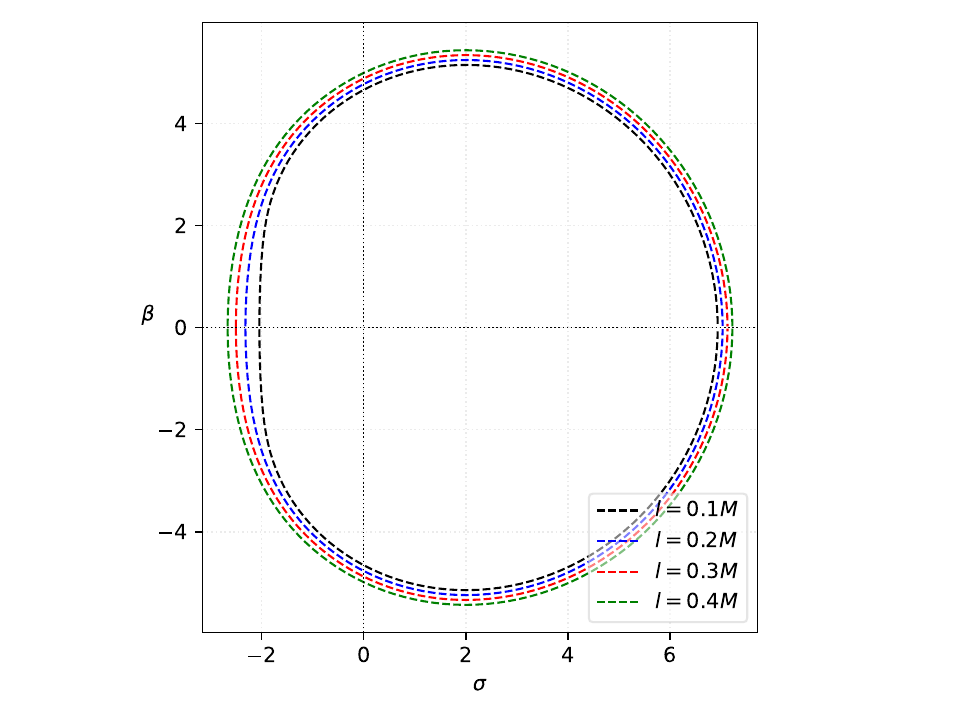}
\includegraphics[width=0.48\textwidth]{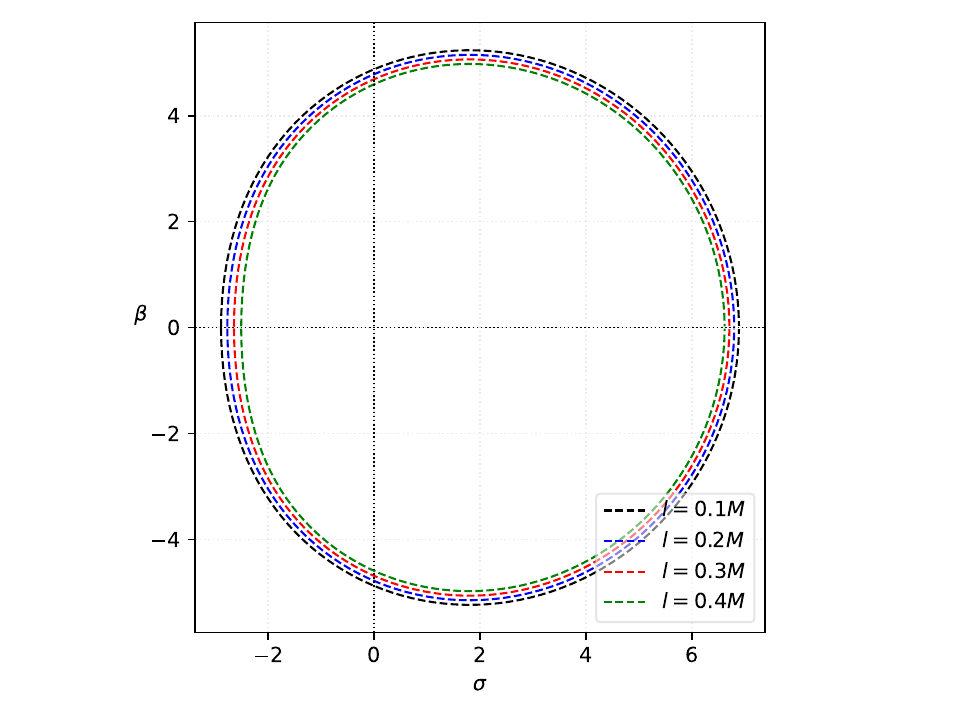}
\includegraphics[width=0.48\textwidth]{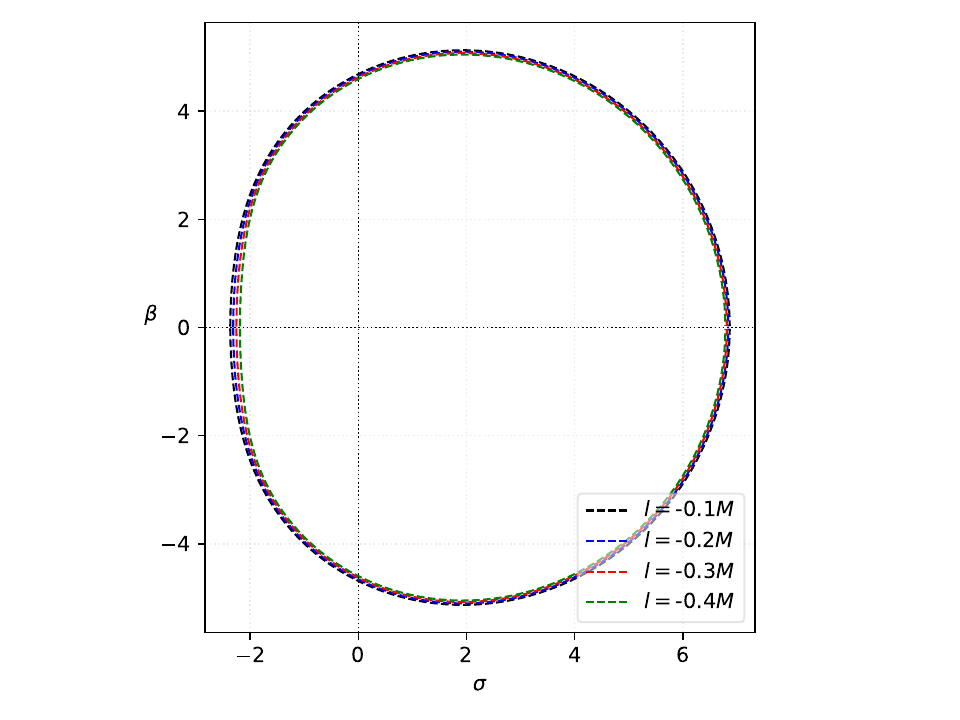}
\includegraphics[width=0.48\textwidth]{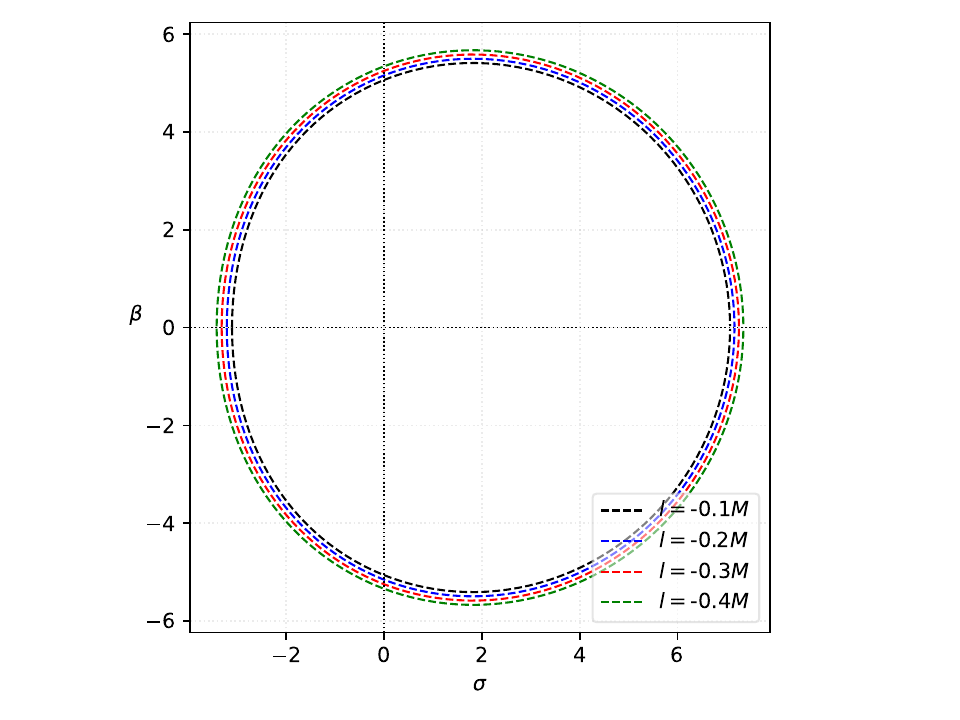}
    \caption{Hairy Kerr black hole shadow ($a = 0.95M$, $\theta = \pi/2$). Upper left panel: $\alpha = 0.35$. Upper right panel: $\alpha = -0.35$. Lower left panel: $\alpha = 0.10$. Lower right panel: $\alpha = -0.35$.}
\end{figure}

The hairy Kerr spacetime by gravitational decoupling has been found in the paper~\cite{bib:rotating}. The $\gamma$ and $M(r$ for this spacetime are given by
\begin{eqnarray} \label{e21}
M(r)&\equiv &M=\text{const.},\nonumber \\
\alpha \gamma(r)&=& \ln \left|1-\alpha \frac{l-re^{-\frac{r}{M}}}{r-2M} \right|,
\end{eqnarray}
The photonsphere radius $r_{\rm ph}$ is found numerically by $\beta=0$ and then
from Eq. \ref{shadowradius}, shadow radius is obtained as shown in Table \ref{tab:Tab1} and in Fig. \ref{hairfig}. Clearly, increasing the parameter l results in an increase in both the photon sphere size and the shadow radius.

\begin{table}[]
\begin{tabular}{|l|l|l|l|}
\hline
      & $r_{\mathrm{ph}}^{-}$ & $r_{\mathrm{ph}}^{+}$ & $R_{\mathrm{sh}}$ \\ \hline
l=0.1 & 2.85   & 3.09   & 5.28   \\ \hline
l=0.3 & 2.89   & 3.12   & 5.37   \\ \hline
l=0.5 & 2.91   & 3.16   & 5.64   \\ \hline
l=0.7 & 2.93   & 3.20   & 6.16   \\ \hline
l=0.9 & 2.95   & 3.26   & 7.16   \\ \hline
\end{tabular}
\caption{Table illustrating the behavior of photon spheres and shadow radius with varying values of the parameter l with constants $M=1$, $\alpha=0.1$, and $a=0.1$.}
  \label{tab:Tab1}
\end{table}

\subsection{Example 2: Rotating Kiselev solution}
\begin{figure} \label{kiselevfig}
    \centering
    \includegraphics[width=0.48\textwidth]{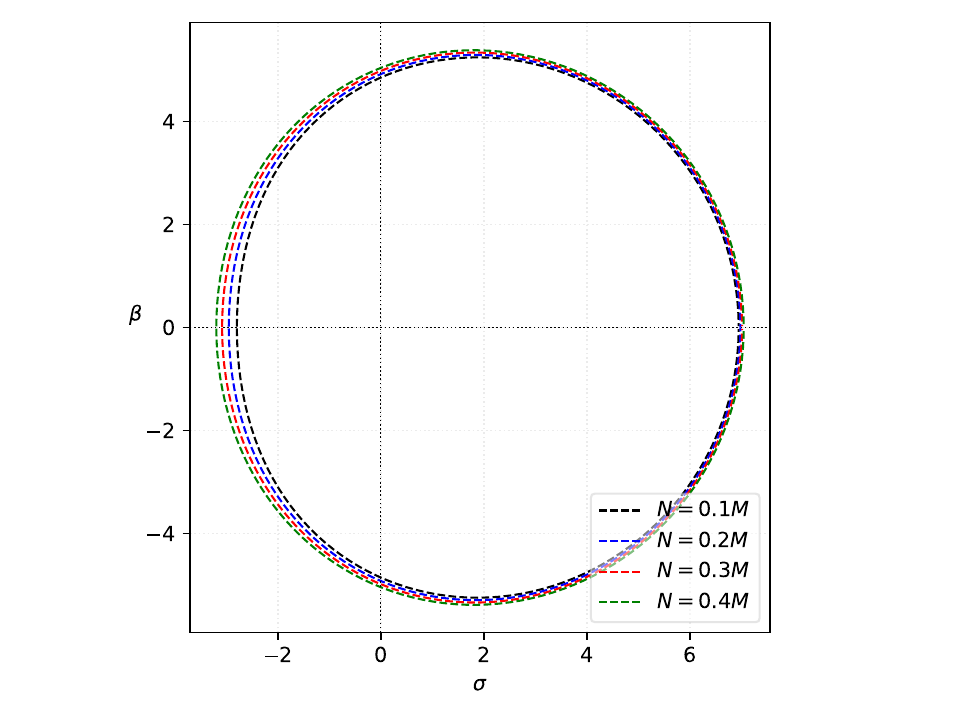}
    \includegraphics[width=0.48\textwidth]{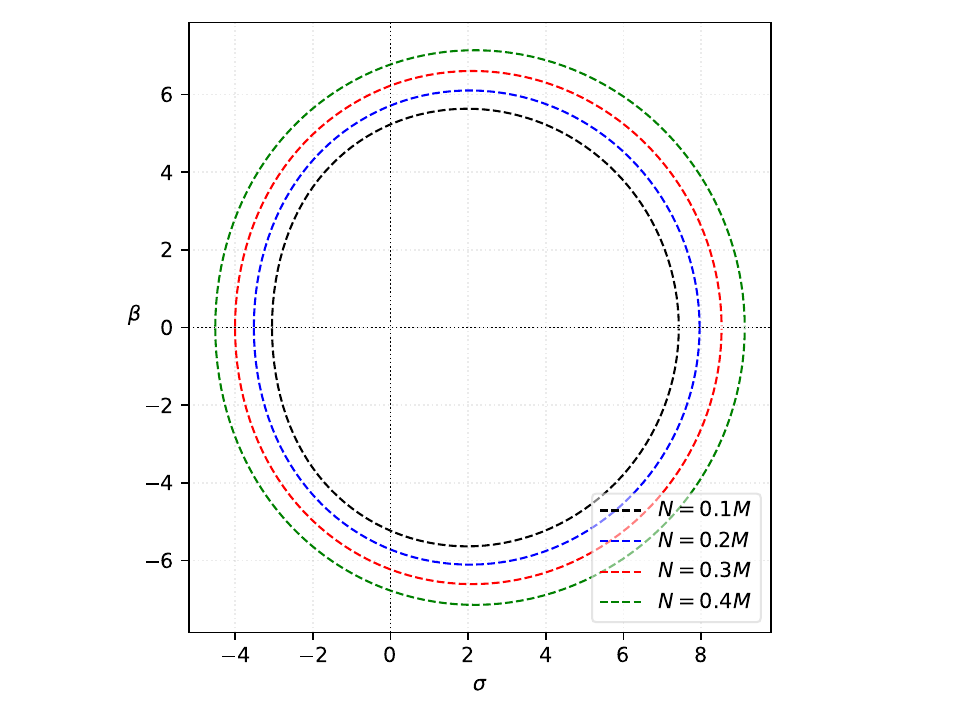}
    \includegraphics[width=0.48\textwidth]{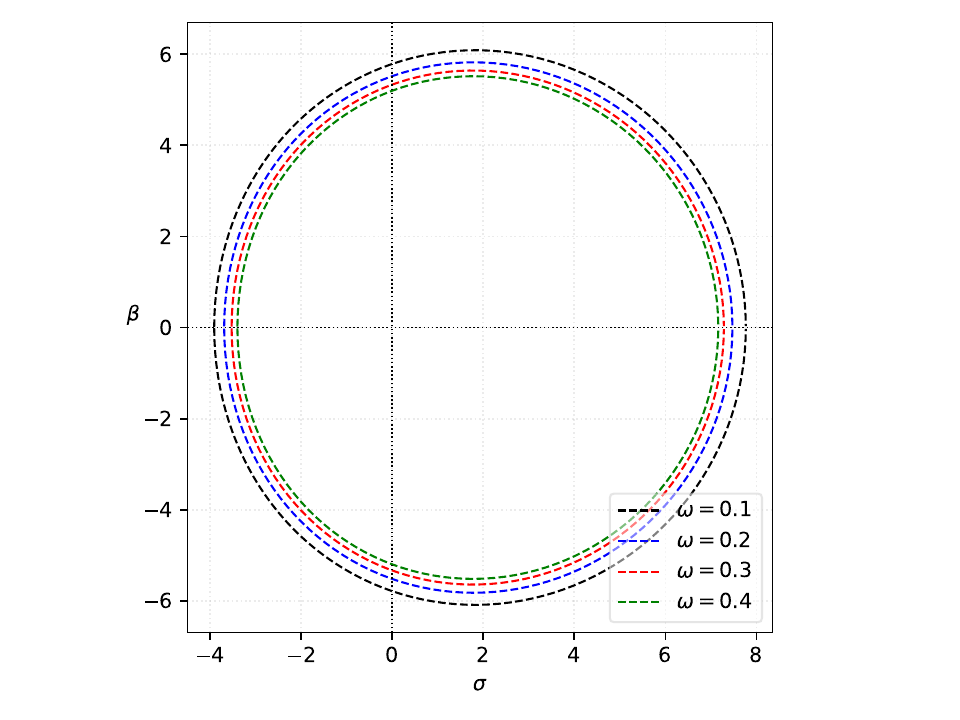}
    \includegraphics[width=0.48\textwidth]{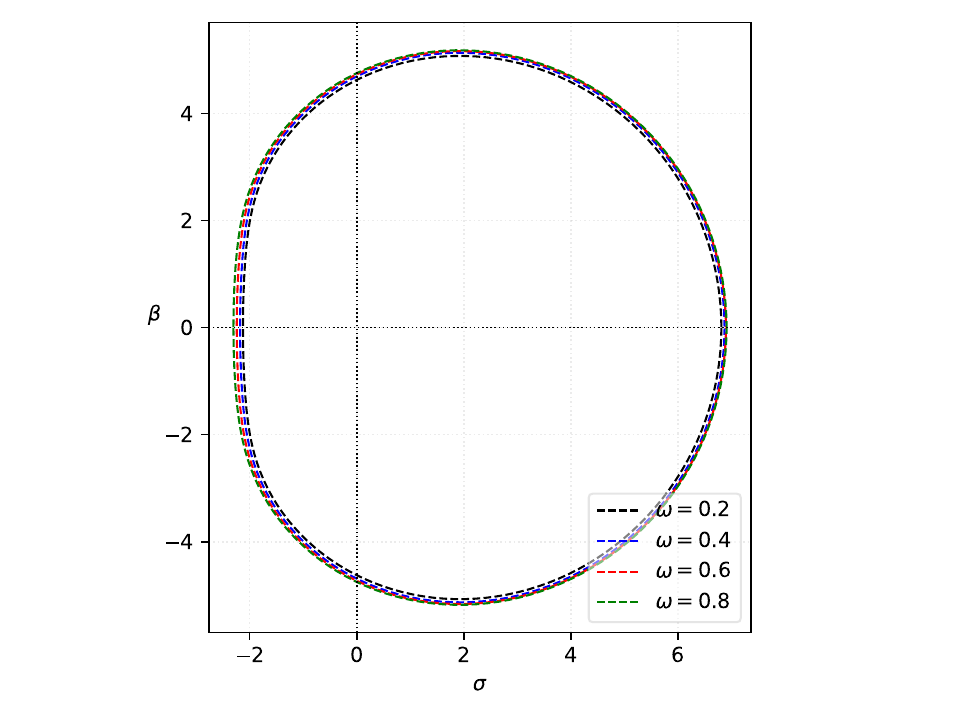}
    \caption{Kiselev shadow ($a = 0.95M$, $\theta = \pi/2$). Upper left panel: $\omega = 0.5$. Upper right panel: $\omega = -0.15$. Lower left panel: $N = 0.5M$. Lower right panel: $N = -0.09M$.}
\end{figure}

In rotating Kiselev solution~\cite{bib:kiselev}, we have
\begin{eqnarray}
M&=&\text{const}.,\nonumber \\
\alpha \gamma(r)&=&\ln \left| 1-\frac{N}{r^{3\omega}(r-2M)}\right|.
\end{eqnarray}
Here $N$ is cosmological parameter and $\omega$ is a parameter of the equation of state. In order to satisfy energy condition $N$ and $\omega$ should have opposite signs.  The photonsphere radius $r_{\rm ph}$ is found numerically by $\beta=0$ and then
from Eq. \ref{shadowradius}, shadow radius is obtained as shown in Table \ref{tab:Tab2} and in Fig. \ref{kiselevfig}. Clearly, increasing the parameter $\omega$ results in an decrease in both the photon sphere size and the shadow radius.
\begin{table}[]
\begin{tabular}{|l|l|l|l|}
\hline
               & $r_{\mathrm{ph}}^{-}$ & $r_{\mathrm{ph}}^{+}$ & $R_{\mathrm{sh}}$ \\ \hline
$\omega$=0.1 & 2.894  & 3.125  & 5.240  \\ \hline
$\omega$=0.2 & 2.891  & 3.123  & 5.224  \\ \hline
$\omega$=0.3 & 2.889  & 3.121  & 5.213  \\ \hline
$\omega$=0.4 & 2.887  & 3.118  & 5.206  \\ \hline
$\omega$=0.5 & 2.886  & 3.117  & 5.202  \\ \hline
\end{tabular}
\caption{Table illustrating the behavior of photon spheres and shadow radius with varying values of the parameter $\omega$  with constants $M=1$, $N=0.01$, $\alpha=0.1$, and $a=0.1$..}
  \label{tab:Tab2}
\end{table}

\subsection{Example 3: Kerr spacetime}

Our last example is the Kerr spacetime, wherein:
\begin{equation} \label{eq:examplern}
\alpha \gamma(r)=\ln \left[\frac{2 M r-r^2}{r (2 M-r)}\right].
\end{equation}
The rotating spacetime also possesses an event horizon, located at $r=r_+$ such that $g^{rr}\left(r_+\right) =0 $, which turns out to be equivalent to the condition 
\begin{equation}
    g_{tt} g_{\phi \phi} - g_{t \phi}^{2} = 0. 
\end{equation}
Solving this equation using the location of the horizon in the Kerr metric and with the ansatz $r_{+}^{\alpha'} = r_{+}^{SK} + \alpha \mu_{1} $, we found that the location of the event horizon for this spacetime is
\begin{equation}
    r_{+} = 2M - \frac{a^2}{2M}  + \frac{\alpha}{6M} + \mathcal{O}(a^2 \alpha, \alpha^{2}).
\end{equation}
Using the condition $g_{tt}(r) = 0$ to find the location of the ergosphere, one finds
\begin{equation}
    r_{erg}^{\alpha'} = 2M - \frac{a^2}{2M} \cos^{2} \theta - \dfrac{\alpha}{12M^{2}} + \mathcal{O}(a^2 \alpha, \alpha^2).
\end{equation}

Note that the $\alpha^\prime$-corrected ergosphere is smaller than the ergosphere of Kerr, namely  $r^{\alpha^\prime}_{\mathrm{erg}}<r_{\mathrm{erg}}^{SK}$, while the location of the event horizon of the $\alpha^\prime$-corrected geometry is larger than the location of the Kerr event horizon

\begin{figure}
    \centering
\label{Kerrfig}
\includegraphics[width=0.48\textwidth]{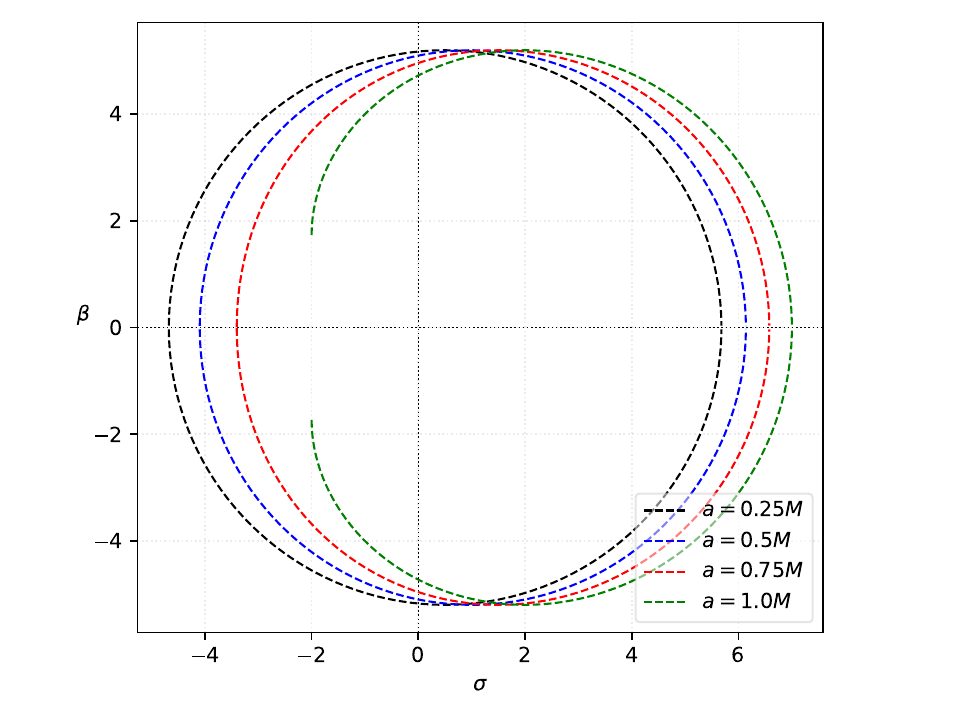}
    \includegraphics[width=0.48\textwidth]{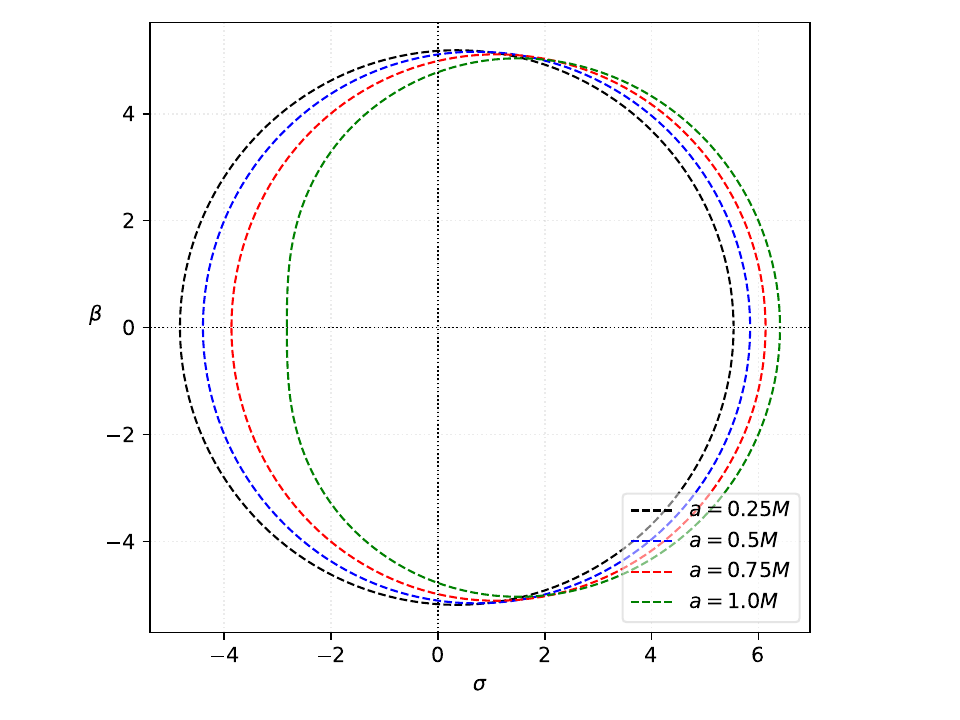}
    \caption{Kerr shadow. Left panel: $\theta = \pi/2$. Right panel: $\theta=\pi/4$.}
\end{figure}

The photonsphere radius $r_{\rm ph}$ is found numerically by $\beta=0$ and then from Eq. \ref{shadowradius}, shadow radius is obtained as shown in Table \ref{tab:Tab3} and in Fig.\ref{Kerrfig}. Clearly, increasing the parameter $M$ results in an inreases in both the photon sphere size and the shadow radius.

\begin{table}[]
\begin{tabular}{|l|l|l|l|}
\hline
  & $r_{\mathrm{ph}}^{-}$ & $r_{\mathrm{ph}}^{+}$ & $R_{\mathrm{sh}}$ \\ \hline
$M$=1.0 & 2.882  & 3.113  & 5.193  \\ \hline
$M$=1.2 & 3.483  & 3.714  & 6.233  \\ \hline
$M$=1.4 & 4.083  & 4.314  & 7.273  \\ \hline
$M$=1.6 & 4.683  & 4.914 & 8.312 \\ \hline
$M$=1.8 & 5.283  & 5.514  & 9.352  \\ \hline
\end{tabular}
\caption{Table illustrating the behavior of photon spheres and shadow radius with varying values of the parameter $M$  with constants $\alpha=0.1$, and $a=0.1$.}
  \label{tab:Tab3}
\end{table}

\section{Conclusions}
A black hole's shadow and photon sphere are critical tools for understanding its properties. Recent observations by the Event Horizon Telescope Collaboration have confirmed that real astrophysical black holes are likely rotating, with a spin of around $0.7$. These black holes are typically surrounded by an accretion disc, which affects the visible size of their shadows. Light propagation in Kerr spacetime, accounting for light deflection due to plasma presence, offers insights into these phenomena. By treating the black hole and accretion disc as a single system, we can better understand light propagation in their combined spacetime.

Gravitational decoupling is a powerful method for analyzing these systems. Our recent work on spherically-symmetric spacetimes has provided clear insights into how geometrical deformations impact black hole shadows and the radius of photon spheres. Building on this, our current study focuses on axially-symmetric spacetimes, where we have found that the influence of geometrical deformations on the apparent size of a black hole image is more intricate.

In this paper, we have analyzed several well-known models and calculated the shadows in these spacetimes using a novel method. Comparing our results with those of an undistorted Kerr spacetime, we have shown that the influence of an accretion disc on black hole geometry is likely negligible. Our approach accurately describes the disc's influence on the apparent shape of the black hole image, providing valuable insights into these complex astrophysical systems.

\acknowledgements
V. Vertogradov thanks the Basis
Foundation (grant number 23-1-3-33-1) for the financial support. A. {\"O}. and R. P. would like to acknowledge networking support by the COST Action CA18108 - Quantum gravity phenomenology in the multi-messenger approach (QG-MM). A. {\"O}. would like to acknowledge networking support by the CA22113 - Fundamental challenges in theoretical physics (THEORY-CHALLENGES) and by the COST Action CA21106 - COSMIC WISPers in the Dark Universe: Theory, astrophysics and experiments (CosmicWISPers).We also thank TUBITAK and SCOAP3 for their support.

\end{document}